\documentclass[12pt]{article}
\usepackage[T1]{fontenc}
\usepackage[utf8]{inputenc}
\usepackage{graphicx}
\usepackage{amsmath}
\usepackage{amssymb}
\usepackage{indentfirst}
\usepackage{latexsym}
\usepackage{multicol}
\usepackage{geometry}
\usepackage{verbatim}
\usepackage{setspace}
\usepackage{float}
\usepackage{siunitx}
\usepackage{listings}
\usepackage{appendix}
\usepackage{xcolor}
\usepackage{caption}
\usepackage{subcaption}
\usepackage{tikz}
\usepackage{ulem}
\usepackage{physics}
\usetikzlibrary{matrix}
\usetikzlibrary{decorations.markings}

\newgeometry{tmargin=2cm, bmargin=2cm, lmargin=2cm, rmargin=2cm}

\begin{document}

\title{
On the non-integrability of three dimensional Ising model
}
\author{Wojciech Niedzi\'o\l ka and Jacek Wojtkiewicz,\\
{\it Faculty of Physics, University of Warsaw,}\\
{\it Pasteura 5, 02-093 Warszawa, Poland}\\
e-mails: ${\rm w.niedziolk2@student.uw.edu.pl}$ (W. N.),
${\rm  wjacek@fuw.edu.pl}$ (J. W.) 
}
\date{}
\maketitle
\abstract{
It is well known that the partition function of two-dimensional Ising model can be expressed
as a Grassmann integral over the action bilinear in Grassmann variables. The key aspect of the
proof of this equivalence is to show that all polygons, appearing in Grassmann integration, enter
with fixed sign. For three-dimensional model, the partition function can also be expressed by
Grassmann integral. However, the action resulting from low-temperature expansion contains
quartic terms, which does not allow explicit computation of the integral. We wanted to check -
apparently not explored - the possibility that using the high-temperature expansion would result in action with only bilinear terms. (in two dimensions, low-T and high-T expansions are equivalent, but in
three dimensions, they differ.) It turned out, however that polygons obtained by Grassman
integration are not of fixed sign for any ordering of Grassmann variables on sites. This way, it
is not possible to express the partition function of three-dimensional Ising model as a Grassman
integral over bilinear action.
}
\vskip0.3cm

\noindent{\it Keywords}: Ising model, Grassmann integration, integrability

\section{Motivation}
The first solution for the two-dimensional {\it Ising model} \cite{Lenz}, \cite{Ising} without a magnetic 
field was obtained by Onsager (1944) \cite{Onsager}. After him, many 
others provided alternative derivations for the free energy, including 
%B. Kaufman
\cite{Kaufman}, 
%M. Kac and J. C. Ward
\cite{Kac&Ward},
\cite{LSM}, 
\cite{Vdv},
% R. J. Baxter and I. G. Enting
\cite{Baxter&Enting}. 
The solution for the three-dimensional model remains an open problem, 
despite numerous attempts and studies that delve into the problem and 
provide new perspectives \cite{Rychkov}. 
There are 
also works by individuals claiming they have solved the problem; however, 
these solutions have not been recognized by the scientific community as 
correct. An example of such an attempt is the work by D. Zhang \cite{Zhang}, 
whose result was not acknowledged due to factors such as the 
dependence of the free energy on boundary conditions in the thermodynamic 
limit \cite{Zhang-com}, \cite{Zhang-com-com}, \cite{Viswanathan}.

In our paper, we wanted to answer the question of why the method of Grassmann
integrals, which allows one to easily obtain an expression for free energy in 2D, fails in 3D. More concretely: In 2D case one can write
the partition function with the use of H-T expansion \cite{Giuliani} as a sum
over closed polygons. Then one can prove that 
the H-T expansion can be obtained as a Grassmann integral with the use of
very natural `action', being the bilinear form over Grassmann variables. Passing to 3D,
it is also possible to obtain the partition function with the use of H-T expansion as a sum
over closed polygons. It is tempting to take the action in the form analogous as in 2D.
But the resulting Grassmann integrals gives apparently wrong expression for the
partition function. We wanted to take a closer look at why it is so. The short summary of the
answer is that in 3D, the Grassmann integral produces certain polygons with a {\it negative sign}.

Let us take a closer look at the utilization of Grassmann variables to solve the Ising model(s) \cite{Samuel1}.
%In this paper, we will focus on modifying S. Samuel's solution for the 
%two-dimensional model \cite{Samuel1}. 
Samuel used Grassmann variables to 
write down the {\it low-temperature expansion} of the Ising Model. His work was 
inspired by the research of C. A. Hurst and H. S. Green, who expressed the 
partition function as a Pfaffian, utilizing a polygonal interpretation of the model
\cite{Hurst&Green}. Hurst and Green referred to their work as a bridge between 
the algebraic approach of Onsager and the combinatorial approach of M. Kac and 
J. C. Ward, which turned out to be more straightforward than both previous works. 
Samuel's 
contribution involved rewriting the expression as an integral over Grassmann 
variables. The Grassmann integral used the {\it quadratic} action, and it was possible
to reproduce Onsager expression for the free energy. In another work, he repeated the 
reasoning in three dimensions \cite{Samuel3}, but the action was {\it quartic} one and 
he wasn't able to obtain an 
explicit expression for the free energy. The modified proof by Samuel 
was presented in A. Giuliani's doctoral thesis \cite{Giuliani}, who reproduced 
Samuel's proof in the language of high-temperature expansion. Giuliani's work 
also proved to be much more transparent (although somewhat lengthy and involved) than the original proof by Samuel, 
which relied on properties of oriented polygons. Giuliani performed an induction 
proof in which, using elementary methods, he proved what Samuel quickly passed 
over by referring to the theorem on oriented polygons.

An important observation is the equivalence between the high-temperature (H-T)
and low-temperature (L-T) expansions in two dimensions. Both expansions are 
described as a sum over polygons on the square lattice. But in H-T expansion, 
polygons are drawn on the original lattice, whereas in L-T case, they are drawn on a {\it dual} one. In 2D, the square lattice is self-dual.
However, it is not the case in 3D. The high-temperature expansion is 
still related to polygons on the lattice, while the low-temperature 
expansion is linked to polyhedrons. Samuel used the low-temperature 
expansion in his works \cite{Samuel1}, \cite{Samuel3}, which did not 
play a significant role in the two-dimensional model. 
In the three-dimensional model, since both expansions are significantly 
different, both formulations need to be examined. Our paper uses
the high-temperature expansion, whereas Samuel studied the three-dimensional 
model using the low-temperature expansion. 

The structure of the paper is as follows. In Sec.~\ref{sec:2D}, we present a sketch of the
solution of the 2D Ising model with the use of Grassmann integrals. This sketch is based on
Giulani paper \cite{Giuliani}, but we also introduce notation that will be used later to simplify formulas. In Sec.~\ref{sec:3D}, we present an attempt to repeat analogous considerations
for 3D model and present points where such an approach fails. The Sec.~\ref{sec:Wnioski} is devoted to summary and conclusions. 

%\section{Ising model's solution using Grassmann variables}

%%%%%%%%%%%%%%%%%%%%%%%%%%%%%%%%%
\section{Two-dimensional model: solution using Grassmann variables}
\label{sec:2D}

We consider classical Ising model with nearest neighbour interaction 
with coupling constant $J_{i j}$ equal to $J_1$ and $J_2$ respectively 
for horizontal and vertical neighbours and following Hamiltonian
\begin{align*}
    \displaystyle
    H(\lbrace s_i \rbrace) &= - \sum_{i j}J_{i j} s_i s_j - \sum_{i}B_i s_i
\end{align*}

We rewrite partition function using high-temperature expansion

\begin{align*}
    \displaystyle
    Z_N &=  \cosh^{M_1}{K_1} \cosh^{M_2}{K_2} 2^N \sum\limits_P t_1^{r_1} t_2^{r_2}
\end{align*}

where $N$ is the number of sites, $M_i$ is the number of 
bonds in each respective direction, $K_{i} = J_i/kT$, $t_{i} = \tanh{K_{i}}$. 
The summation is over all multipolygons, which are geometrical figures composed 
of bonds on the lattice with an even number of bonds connected to each site. 
$r_i$ represents the number of bonds in each respective direction.

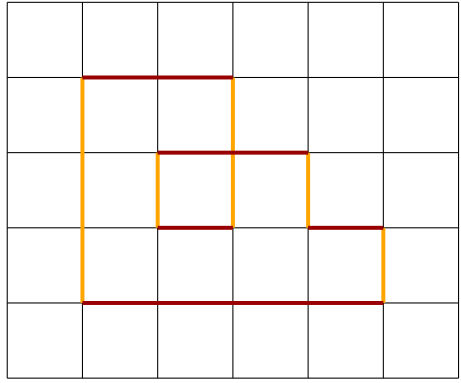
\begin{figure}[H]
    \centering
    \begin{tikzpicture}
        \def\cellsize{1cm}
        \def\hcolor{rgb,1:red,1;green,0.647;blue,0}
        \def\vcolor{rgb,1:red,0.6;green,0;blue,0}
        % Draw the horizontal lines
        \foreach \y in {0,...,5} {
            \draw (0,\y*\cellsize) -- (6*\cellsize,\y*\cellsize);
        }
        % Draw the vertical lines
        \foreach \x in {0,...,6} {
            \draw (\x*\cellsize,0) -- (\x*\cellsize,5*\cellsize);
        }
        \draw[color = \hcolor, line width = 1.5] (1,1) -- (1,4);
        \draw[color = \hcolor, line width = 1.5] (3,2) -- (3,4);
        \draw[color = \hcolor, line width = 1.5] (4,3) -- (4,2);
        \draw[color = \hcolor, line width = 1.5] (2,3) -- (2,2);
        \draw[color = \hcolor, line width = 1.5] (5,1) -- (5,2);
        \draw[color = \vcolor, line width = 1.5] (1,1) -- (5,1);
        \draw[color = \vcolor, line width = 1.5] (1,4) -- (3,4);
        \draw[color = \vcolor, line width = 1.5] (2,3) -- (4,3);
        \draw[color = \vcolor, line width = 1.5] (2,2) -- (3,2);
        \draw[color = \vcolor, line width = 1.5] (4,2) -- (5,2);
    \end{tikzpicture}
    \caption{An example of multipolygon with 10 horizontal edges with constant {\textcolor[rgb]{0.6,0,0}{$t_1$}} and 
    8 vertical ones with constant {\textcolor[rgb]{1,0.647,0}{$t_2$}}}
\end{figure}

Samuel noticed \cite{Samuel1} that it is possible to 
relate a statistical sum to an integral over Grassmann variables. 
His reasoning reproduced Onsager's result and led to a new method 
for solving the two-dimensional Ising model. Below, we present a sketch 
of the reasoning, which serves as a starting point for considering the 
three-dimensional case. However, it should be noted that the conventions 
used in this work are taken from Giuliani's paper \cite{Giuliani}, where 
he reproduced Samuel's proof in two dimensions using the high-temperature expansion.

Let us consider $S$, which is a bilinear action in Grassmann variables:

\begin{align}
    \displaystyle
    S = \sum\limits_{i}{}' t_1 \overline{H}_i H_{i+\hat{e}_1} + t_2 \overline{V}_i V_{i+\hat{e}_2} +
    \sum\limits_i \overline{H}_i H_i + \overline{V}_i V_{i} + \overline{V}_i \overline{H}_i + 
    V_i \overline{H}_i + H_i \overline{V}_i + V_i H_i
\end{align}

where $\hat{e}_1$ and $\hat{e}_2$ are the coordinate versors in the horizontal 
and vertical directions, respectively and the primed sum runs over those 
sites for which $i$, $i+\hat{e}_1$, 
and $i+\hat{e}_2$ belong to the lattice. This form of $S$ corresponds to 
open boundary conditions. Then the sum over polygons for open boundary 
conditions turns out to be equal to:

\begin{align}
    \displaystyle
    \sum\limits_P t_1^{r_1} t_2^{r_2} = (-1)^{N} \int \prod_{i} d \overline{H}_i dH_i d \overline{V}_i dV_i e^S
\end{align}

By modifying the action, one can also obtain a formula for periodic 
boundary conditions. Diagonalizing the functional and then using the 
properties of Grassmann integrals, one can derive the formula for the 
free energy obtained by Onsager.

Let us examine the method of calculating this integral in the two-dimensional 
case. Due to the properties of Grassmann variables, non-zero contributions 
come only from terms containing all 4N variables exactly once. Each of 
these terms can be represented as a so-called dimer on a new lattice 
obtained by replacing the sitess of the original lattice with four sites 
on the new lattice, as shown in Figure\ref{wieldim}. While for the square 
correspondence is one to one, for a general multipolygon we might have 
many different dimers corresponding to it, as we have 3 different ways of 
representing an empty site.
The key to proving the relationship between the integral and the statistical 
sum is to show that for each multipolygon, all dimers corresponding to it 
together give a contribution of $(-1)^{n_\gamma} t_1^{r_1} t_2^{r_2}$, not 
including empty sites. 
This weight is composed of the constants $t_i$ appearing in 
the appropriate powers related to the number of bonds in both directions, 
and an additional factor called the sign of the polygon.
Here, $n_\gamma$ denotes the number of sites belonging to the multipolygon.

Additionally, the integral for a so-called empty site, i.e., a site 
through which no polygon passes, which is the sum of three possible 
configurations of variables at that site, gives -1, which can be easily verified:

\begin{align*}
    \int d \overline{H} dH d \overline{V} dV (\overline{H} H \overline{V} V+ 
    \overline{V} \overline{H} V H + V \overline{H} H \overline{V}) = 1 - 1 - 1 = -1
\end{align*}

Thus, the contribution from the entire lattice for a single 
multipolygon, including empty sites, is 
$(-1)^{n_\gamma'} (-1)^{n_\gamma} t_1^{r_1} t_2^{r_2} = (-1)^{N} t_1^{r_1} t_2^{r_2}$, 
$n_\gamma'$ is the number of empty sites, here we used the fact that the total number of 
sites is $N$. 
By summing 
over all configurations of polygons, we obtain the above formula.

Here, we would like to introduce a notation for pairs of variables 
$\overline{H}_i$, $H_i$ and $\overline{V}_i$, $V_i$. They are of 
the same type (either H or V), while the pairs $V_i$, $H_i$ and 
$\overline{V}_i$, $\overline{H}_i$ are from the same group (group 1 and 2, respectively). 
These variables anti-commute, which means that their pairs commute. 
Therefore, the pairs of variables can be rearranged without changing 
the result under the integral. According to the following rule, we 
choose a direction for the polygon and one of the edges connecting 
the vertices, and write down its term. Then, we proceed to the next 
points in the chosen order, writing down the term associated with 
connecting two variables at each point, followed by the term associated 
with connecting that point with the next one. For the terms connecting 
the vertices, if the order of points corresponding to variables in the 
functional is different from the one resulting from the direction of the 
contour, we change the order of variables, obtaining either 
$-H_{i+\hat{e}_1} \overline{H}_i$ or $-V_{i+\hat{e}_1} \overline{V}_i$. 
The minus sign is "attached" to the variable from group 2. Finally, we 
only need to move the first variable from the beginning to the end in 
order to connect it with the remaining variables belonging to the same 
point. This introduces an additional minus sign since, apart from the 
first variable, there is an odd number of variables. The procedure described 
above, for an example polygon in the form of a square, looks as follows:

\begin{figure}[H]
    \begin{subfigure}[h]{0.3\linewidth}
        \begin{tikzpicture}[scale=2.5, decoration={markings, mark=at position 0.5 with {\arrow{>}}}    ]
            \coordinate (A) at (0,0);
            \coordinate (B) at (1,0);
            \coordinate (C) at (1,1);
            \coordinate (D) at (0,1);
            % Draw edges
            \draw[postaction={decorate},line width=1.5pt] (A) -- (B);
            \draw[postaction={decorate},line width=1.5pt] (B) -- (C);
            \draw[postaction={decorate},line width=1.5pt] (C) -- (D);
            \draw[postaction={decorate},line width=1.5pt] (D) -- (A);
            % Label vertices
            \node at (A) [below left] {$A$};
            \node at (B) [below right] {$B$};
            \node at (C) [above right] {$C$};
            \node at (D) [above left] {$D$};
            \end{tikzpicture}
            \subcaption{oriented polygon}
    \end{subfigure}
    \begin{subfigure}[h]{0.3\linewidth}
        \begin{tikzpicture}[scale=2.5, decoration={markings, mark=at position 0.5 with {\arrow{>}}}    ]
            \coordinate (A1) at (0.1,0);
            \coordinate (B1) at (1.1,0);
            \coordinate (C1) at (1.1,1);
            \coordinate (D1) at (0.1,1);
            \coordinate (A2) at (0,0.1);
            \coordinate (B2) at (1,0.1);
            \coordinate (C2) at (1,1.1);
            \coordinate (D2) at (0,1.1);
            \coordinate (A3) at (-0.1,0);
            \coordinate (B3) at (0.9,0);
            \coordinate (C3) at (0.9,1);
            \coordinate (D3) at (-0.1,1);
            \coordinate (A4) at (0,-0.1);
            \coordinate (B4) at (1,-0.1);
            \coordinate (C4) at (1,0.9);
            \coordinate (D4) at (0,0.9);
            \foreach \pt in {A1,B1,C1,D1,A2,B2,C2,D2,A3,B3,C3,D3,A4,B4,C4,D4}
                \draw (\pt) node[circle, fill, inner sep=1pt] {};
            \draw (A1) -- (B3);
            \draw (B2) -- (C4);
            \draw (C3) -- (D1);
            \draw (D4) -- (A2);
            \draw (A4) -- (A3);
            \draw (B1) -- (B4);
            \draw (C1) -- (C2);
            \draw (D2) -- (D3);
            % Label vertices
            \end{tikzpicture}
            \subcaption{dimer corresponding to a square}
    \end{subfigure}
    \begin{subfigure}[h]{0.3\linewidth}
        \begin{tikzpicture}[scale = 1.5]
            \coordinate (origin) at (0, 0);
            \coordinate (X1) at (1, 0);
            \coordinate (X2) at (-1, 0);
            \coordinate (Z1) at (0, 1);
            \coordinate (Z2) at (0, -1);
            \foreach \pt in {X1, X2, Z1, Z2}
                \draw[dashed] (origin) -- (\pt);
            \foreach \pt in {X1, X2, Z1, Z2}
                \draw (\pt) node[circle, fill, inner sep=1pt] {};
            \node at (X1) [right] {$\overline{H}$};
            \node at (X2) [left] {$H$};
            \node at (Z1) [above] {$\overline{V}$};
            \node at (Z2) [below] {$V$};
        \end{tikzpicture}
        \subcaption{Grassmann variables associated to a site}
    \end{subfigure}
    \caption{An example of a polygon and corresponding dimmer}
    \label{wieldim}
\end{figure}
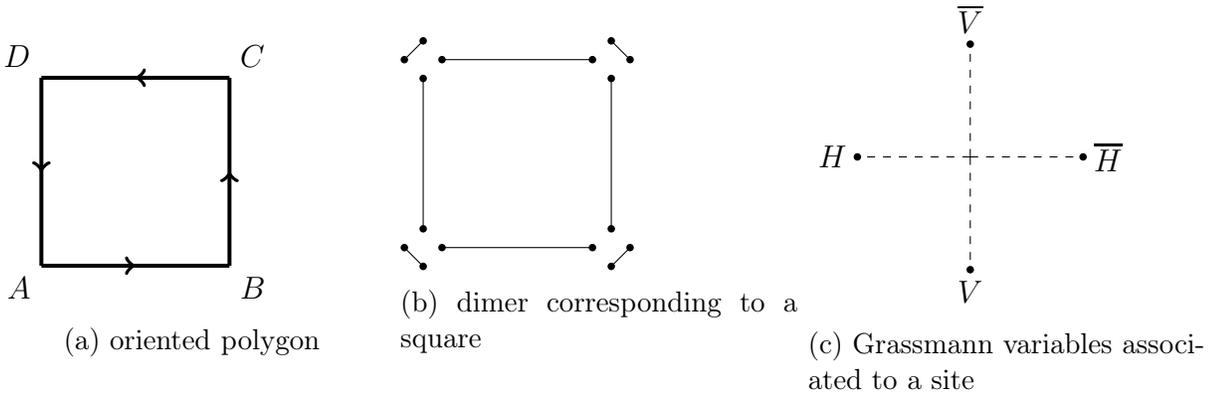

\begin{align*}
    \displaystyle
    \int \prod_{i} d \overline{H}_i dH_i d \overline{V}_i dV_i 
    \overline{V}_A V_D \cdot H_A V_A \cdot \overline{H}_A H_B \cdot
    V_B \overline{H}_B \cdot \overline{V}_B V_C \cdot \overline{V}_C 
    \overline{H}_C \cdot \overline{H}_D H_C \cdot V_D \overline{H}_D = \\
    \int \prod_{i} d \overline{H}_i dH_i d \overline{V}_i dV_i 
    V_D (-\overline{V}_A) \cdot H_A V_A \cdot \overline{H}_A H_B \cdot
    V_B \overline{H}_B \cdot \overline{V}_B V_C \cdot \overline{V}_C 
    \overline{H}_C \cdot  H_C (-\overline{H}_D) \cdot V_D \overline{H}_D = \\
    - \int \prod_{i} d \overline{H}_i dH_i d \overline{V}_i dV_i 
    (-\overline{V}_A)  H_A V_A  \overline{H}_A \cdot H_B 
    V_B \overline{H}_B  \overline{V}_B \cdot V_C  \overline{V}_C 
    \overline{H}_C   H_C \cdot (-\overline{H}_D)  V_D \overline{H}_D V_D
\end{align*}

Performing the integration for each point, 
it can be seen that sign of the square is correct and 
equal to +1. As shown in the example, this procedure 
allows us to express the sign of a polygon as (-1) times 
the product of integrals for individual points. Additionally, 
the integral for a given node depends solely on through which 
variable we enter the site and through which we leave it while 
traversing the polygon in a predetermined direction. Let us 
define an auxiliary function:

\begin{align}
    F_2(X, Y) = \int d \overline{H} dH d \overline{V} dV (\pm X \dots Y)
\end{align}

where X and Y are two distinct Grassmann variables, and 
the dots represent the remaining two variables in the 
order determined by the action. The sign is + if X 
belongs to group 1 and - if it belongs to group 2. Thus, 
the sign of the square can be written as 
$-F_2(\overline{V}, \overline{H}) F_2(H, \overline{V}) F_2(V, H) F_2(\overline{H}, V)$.

Furthermore, we will need a simple property of $F_2$, namely 
$F_2(X,Y) = \pm F_2(Y,X)$, where + corresponds to X and Y 
being from different groups, and - when they are from the same 
group. This follows from the necessity of performing an odd 
number of variable exchanges to swap X and Y. Thus, after such 
an exchange, the sign under the integral changes, meaning that 
the integral does not depend on the order of X and Y if they 
are from different groups.

%%%%%%%%%%%%%%%%%%%%%%%%%%%%%%
\section{Three-dimensional model}
\label{sec:3D}

Now, we pass to three dimensions. 
Naturally, in order to demonstrate the connection between 
the Grassmann integral and the high-temperature 
representation, we want the integral for any given multipolygon
to yield the corresponding contribution $\alpha t_1^{r_1} t_2^{r_2} t_3^{r_3}$, 
again not counting empty sites. 
Analogously to the two-dimensional case, we may want $\alpha = (-1)^{n_\gamma}$ 
and require the integral for a single point to be -1. However, we will consider 
a more general case.
We desire the sum over multipolygon to be equal to the Grassmann variable 
integral multiplied by a constant A. Upon expanding $e^S$, we obtain 
terms for different configurations that, after integration, result in 
the sum of terms $A w^{N-n_\gamma} \alpha t_1^{r_1} t_2^{r_2} t_3^{r_3}$, 
where $w$ represents the weight of an empty sites. We aim to make the weights 
of the polygons, together with the empty sites and the global constant $A$, 
i.e., $A w^{N-n_\gamma} \alpha$, equal to 1. This gives us a condition that 
$\alpha$ is the same for any two polygon configurations with the same $r_1$, $r_2$, 
and $r_3$, as well as the same number of points belonging to the polygon. 
In the following part, we will show that this condition cannot be satisfied.

To each node, we add additional Grassmann variables $U_i$ and $\overline{U}_i$. 
We extend the notions of type and group to incorporate the new variables, where 
the new variables form a new type U, with the first one joining group 1 and the 
second one joining group 2. However, we need to specify how to extend the action 
with the new variables. Naturally, we want all 15 terms connecting the variables 
at a given point to appear, as well as 3 types of connections between points. 
Hence, the new action takes the form:

\begin{align}
    \displaystyle
    S_{3D} = \sum\limits_{i}{}' \left( t_1 \overline{H}_i H_{i+\hat{e}_1} + t_2 \overline{V}_i V_{i+\hat{e}_2} +
    t_3 \overline{U}_i U_{i+\hat{e}_3} \right) +
    \sum\limits_{i}{} \left( a_1 \overline{H}_i H_i + a_2 \overline{V}_i V_{i} 
    + a_3 \overline{U_i} U_i + \right.\\ + 
    a_4 H_i V_i + a_5 H_i U_i + a_6 H_i \overline{V} +
    a_7 H_i \overline{U_i} + 
    a_8 V_i U_i + a_9 V_i \overline{H_i} + \notag \\ + 
    a_{10} V_i \overline{U_i} + a_{11} U_i \overline{H_i} + 
    a_{12} U_i \overline{V_i} + a_{13} \overline{H_i} \overline{V_i} 
    + a_{14} \overline{H_i} \overline{U_i} + a_{15} \overline{V_i} \overline{U_i}
    \left. \right)\notag
\end{align}

where $a_i$ represents certain constants. 
The natural choice would be $a_i=\pm$, but the final
proof holds true even if we allow $a_i \in \mathbb{C}$.
Analogous to the two-dimensional case, we consider open boundary 
conditions. The summation with a prime runs over those points for 
which i, $i+\hat{e}_1$, $i+\hat{e}_2$, $i+\hat{e}_3$ belong to the 
lattice. It turns out that regardless of how we choose the action, 
certain polygons will always yield opposite signs, thus it is impossible 
to select an action that expresses the statistical sum through the 
Grassmann variable integral.

Now, let's define the function $F_3(X, Y)$ analogously to 
$F_2(X, Y)$. The new function will differ in the sense that 
$\dots$ now contain 4 additional variables, and thus they will 
be the sum of 3 possible products of those variables with the 
new action. Each of these products yields $\pm1$, so the possible 
values of $F_3(X, Y)$ are $\pm1$,$\pm3$. The polygon weights $\alpha$ 
are the product of the $F_3$ functions. In the two-dimensional case, 
since $F_2(X,Y)=\pm1$, we had $\alpha = \pm1$. However, in three dimensions, 
$\alpha$ generally equals $\pm 3^k$ for some natural number k, which means 
there are more possible weights for polygons. For example, $F_3(H, U)$ will 
consist of the following 3 terms corresponding to different dimer configurations 
at the vertex:

\begin{align*}
    F_3(H, U) = &\int d \overline{H} dH d \overline{V} dV d \overline{U} dU \\
    &( H \cdot a_2 \overline{V} V \cdot a_{14}\overline{H} \overline{U}\cdot U +
    H \cdot a_{10} V \overline{U}\cdot a_{13} \overline{H} \overline{V} \cdot U +
    H \cdot a_9 V \overline{H} \cdot a_{15} \overline{V} \overline{U}\cdot U) \\
    = &\int d \overline{H} dH d \overline{V} dV d \overline{U} dU \\
    & H ( a_2 a_{14} \cdot \overline{V} V \cdot \overline{H} \overline{U} +
    a_{10} a_{13} \cdot \overline{U} V \cdot \overline{H} \overline{V} +
    a_9 a_{15} \cdot \overline{H} V \cdot \overline{V} \overline{U}) U
\end{align*}

where $a_i$ are constants associated with the definition 
of the action, specifically the order in which we selected 
the terms for these expressions.

\begin{figure}[H]
    \begin{subfigure}[h]{0.3\linewidth}
        \begin{tikzpicture}[scale = 1.5]
            \coordinate (origin) at (0, 0, 0);
            \coordinate (X1) at (1, 0, 0);
            \coordinate (X2) at (-1, 0, 0);
            \coordinate (Y1) at (0, 0, 1);
            \coordinate (Y2) at (0, 0, -1);
            \coordinate (Z1) at (0, 1, 0);
            \coordinate (Z2) at (0, -1, 0);
            \foreach \pt in {X1, X2, Y1, Y2, Z1, Z2}
                \draw[dashed] (origin) -- (\pt);
            \foreach \pt in {X1, X2, Y1, Y2, Z1, Z2}
                \draw (\pt) node[circle, fill, inner sep=1pt] {};
            \draw (Y1) -- (Y2);
            \draw (X1) -- (Z1);
            \node at (X1) [right] {$\overline{H}$};
            \node at (X2) [left] {$H$};
            \node at (Y1) [below left] {$V$};
            \node at (Y2) [above right] {$\overline{V}$};
            \node at (Z1) [above] {$\overline{U}$};
            \node at (Z2) [below] {$U$};
        \end{tikzpicture}
    \end{subfigure}
    \begin{subfigure}[h]{0.3\linewidth}
        \begin{tikzpicture}[scale = 1.5]
            \coordinate (origin) at (0, 0, 0);
            \coordinate (X1) at (1, 0, 0);
            \coordinate (X2) at (-1, 0, 0);
            \coordinate (Y1) at (0, 0, 1);
            \coordinate (Y2) at (0, 0, -1);
            \coordinate (Z1) at (0, 1, 0);
            \coordinate (Z2) at (0, -1, 0);
            \foreach \pt in {X1, X2, Y1, Y2, Z1, Z2}
                \draw[dashed] (origin) -- (\pt);
            \foreach \pt in {X1, X2, Y1, Y2, Z1, Z2}
                \draw (\pt) node[circle, fill, inner sep=1pt] {};
            \draw (Y1) -- (Z1);
            \draw (X1) -- (Y2);
            \node at (X1) [right] {$\overline{H}$};
            \node at (X2) [left] {$H$};
            \node at (Y1) [below left] {$V$};
            \node at (Y2) [above right] {$\overline{V}$};
            \node at (Z1) [above] {$\overline{U}$};
            \node at (Z2) [below] {$U$};
        \end{tikzpicture}
    \end{subfigure}
    \begin{subfigure}[h]{0.35\linewidth}
        \begin{tikzpicture}[scale = 1.5]
            \coordinate (origin) at (0, 0, 0);
            \coordinate (X1) at (1, 0, 0);
            \coordinate (X2) at (-1, 0, 0);
            \coordinate (Y1) at (0, 0, 1);
            \coordinate (Y2) at (0, 0, -1);
            \coordinate (Z1) at (0, 1, 0);
            \coordinate (Z2) at (0, -1, 0);
            \foreach \pt in {X1, X2, Y1, Y2, Z1, Z2}
                \draw[dashed] (origin) -- (\pt);
            \foreach \pt in {X1, X2, Y1, Y2, Z1, Z2}
                \draw (\pt) node[circle, fill, inner sep=1pt] {};
            \draw (Y1) -- (X1);
            \draw (Y2) -- (Z1);
            \node at (X1) [right] {$\overline{H}$};
            \node at (X2) [left] {$H$};
            \node at (Y1) [below left] {$V$};
            \node at (Y2) [above right] {$\overline{V}$};
            \node at (Z1) [above] {$\overline{U}$};
            \node at (Z2) [below] {$U$};
        \end{tikzpicture}
    \end{subfigure}
    \caption{Various configurations of connections between 
    four variables $V$, $\overline{V}$, $\overline{H}$, $\overline{U}$ 
    corresponding to successive expressions under the integral.
    }
\end{figure}
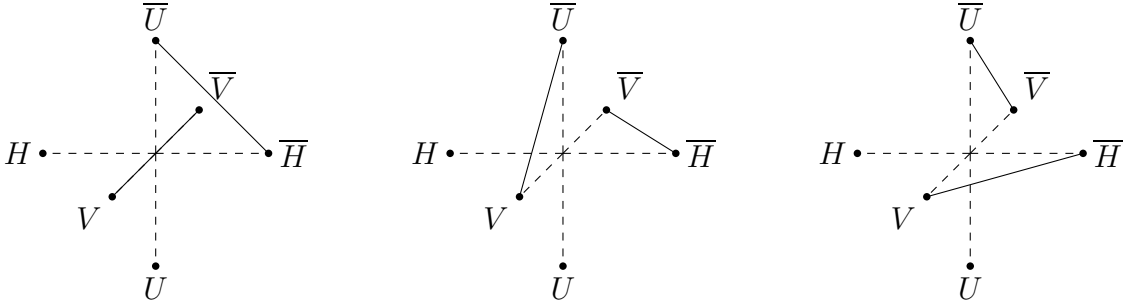

The aim of this study was to demonstrate that no choice of 
$\alpha$ and action can guarantee the condition of $\alpha$ 
depending solely on $r_i$ and $n_\gamma$. The research commenced 
with considering polygons on a 2x2x2 lattice, which is the simplest 
three-dimensional lattice. The analysis focused on the simplest 
non-planar polygons with two edges in each of the three directions. 
Two exemplary polygons are visible in the figure below. The Grassmann 
variables are denoted as $X$, $X'$, $Y$, $Y'$, $Z$, $Z'$, which are, 
of course, the six variables defined earlier, but after a certain rotation.

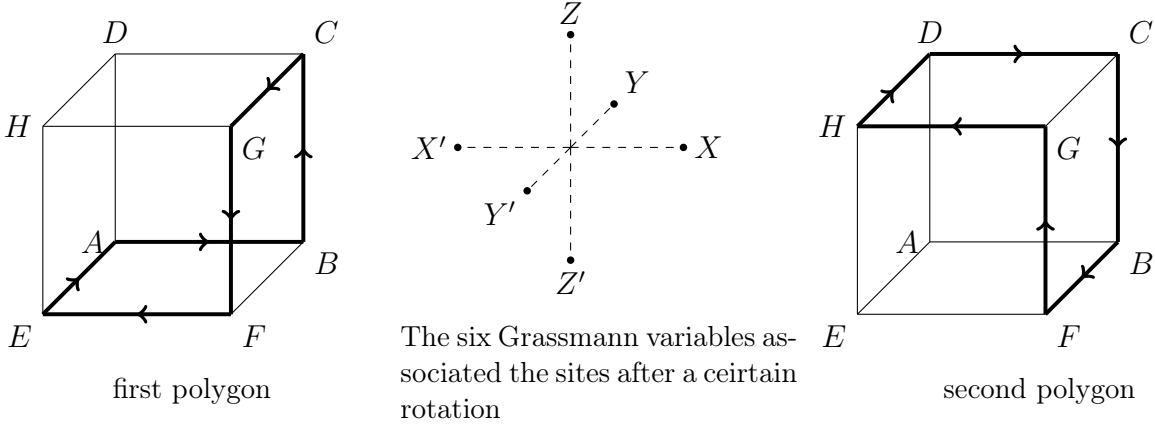
\begin{figure}[H]
    \begin{subfigure}[h]{0.3\linewidth}
        % \begin{tikzpicture}
        %     \matrix (m) [matrix of math nodes, row sep=3em,
        %     column sep=3em]{
        %     & H & & G \\
        %     E & & F & \\
        %     & D & & C \\
        %     A & & B & \\};
        %     \path[-stealth]
        %     (m-1-2) edge[-] (m-1-4) edge[-] (m-2-1) edge[-] (m-3-2)
        %     (m-1-4) edge[->,line width=2pt] (m-3-4) edge[<-,line width=2pt] (m-2-3)
        %     (m-2-1) edge[-] (m-2-3) edge[-] (m-4-1)
        %     (m-2-3) edge[<-,line width=2pt] (m-4-3)
        %     (m-3-2) edge[<-,line width=2pt] (m-3-4) edge[->,line width=2pt] (m-4-1)
        %     (m-3-4) edge[-] (m-4-3)
        %     (m-4-1) edge[->,line width=2pt] (m-4-3);
        % \end{tikzpicture}
        \begin{tikzpicture}[scale=2.5, decoration={markings, mark=at position 0.5 with {\arrow{>}}}    ]
            % Define vertices
            \coordinate (A) at (0,0,0);
            \coordinate (B) at (1,0,0);
            \coordinate (C) at (1,1,0);
            \coordinate (D) at (0,1,0);
            \coordinate (E) at (0,0,1);
            \coordinate (F) at (1,0,1);
            \coordinate (G) at (1,1,1);
            \coordinate (H) at (0,1,1);
            % Draw edges
            \draw[postaction={decorate},line width=1.5pt] (A) -- (B);
            \draw[postaction={decorate},line width=1.5pt] (B) -- (C);
            \draw (C) -- (D);
            \draw (D) -- (A);
            \draw[postaction={decorate},line width=1.5pt] (E) -- (A);
            \draw (B) -- (F);
            \draw[postaction={decorate},line width=1.5pt] (C) -- (G);
            \draw (D) -- (H);
            \draw[postaction={decorate},line width=1.5pt] (F) -- (E);
            \draw[postaction={decorate},line width=1.5pt] (G) -- (F);
            \draw (G) -- (H);
            \draw (H) -- (E);
            % Label vertices
            \node at (A) [left] {$A$};
            \node at (B) [below right] {$B$};
            \node at (C) [above right] {$C$};
            \node at (D) [above] {$D$};
            \node at (E) [below left] {$E$};
            \node at (F) [below right] {$F$};
            \node at (G) [below right] {$G$};
            \node at (H) [left] {$H$};
            \end{tikzpicture}
            \subcaption*{first polygon}
    \end{subfigure}
    \begin{subfigure}[h]{0.3\linewidth}
        \begin{tikzpicture}[scale = 1.5]
            \coordinate (origin) at (0, 0, 0);
            \coordinate (X1) at (1, 0, 0);
            \coordinate (X2) at (-1, 0, 0);
            \coordinate (Y1) at (0, 0, 1);
            \coordinate (Y2) at (0, 0, -1);
            \coordinate (Z1) at (0, 1, 0);
            \coordinate (Z2) at (0, -1, 0);
            \foreach \pt in {X1, X2, Y1, Y2, Z1, Z2}
                \draw[dashed] (origin) -- (\pt);
            \foreach \pt in {X1, X2, Y1, Y2, Z1, Z2}
                \draw (\pt) node[circle, fill, inner sep=1pt] {};
            \node at (X1) [right] {$X$};
            \node at (X2) [left] {$X'$};
            \node at (Y1) [below left] {$Y'$};
            \node at (Y2) [above right] {$Y$};
            \node at (Z1) [above] {$Z$};
            \node at (Z2) [below] {$Z'$};
        \end{tikzpicture}
        \subcaption*{The six Grassmann variables associated the sites after a ceirtain rotation}
    \end{subfigure}
    \begin{subfigure}[h]{0.35\linewidth}
        \begin{tikzpicture}[scale=2.5, decoration={markings, mark=at position 0.5 with {\arrow{>}}}    ]
            % Define vertices
            \coordinate (A) at (0,0,0);
            \coordinate (B) at (1,0,0);
            \coordinate (C) at (1,1,0);
            \coordinate (D) at (0,1,0);
            \coordinate (E) at (0,0,1);
            \coordinate (F) at (1,0,1);
            \coordinate (G) at (1,1,1);
            \coordinate (H) at (0,1,1);
            % Draw edges
            \draw (A) -- (B);
            \draw[postaction={decorate},line width=1.5pt] (C) -- (B);
            \draw[postaction={decorate},line width=1.5pt] (D) -- (C);
            \draw (D) -- (A);
            \draw (E) -- (A);
            \draw[postaction={decorate},line width=1.5pt] (B) -- (F);
            \draw (C) -- (G);
            \draw[postaction={decorate},line width=1.5pt] (H) -- (D);
            \draw (F) -- (E);
            \draw[postaction={decorate},line width=1.5pt] (F) -- (G);
            \draw[postaction={decorate},line width=1.5pt] (G) -- (H);
            \draw (H) -- (E);
            % Label vertices
            \node at (A) [left] {$A$};
            \node at (B) [below right] {$B$};
            \node at (C) [above right] {$C$};
            \node at (D) [above] {$D$};
            \node at (E) [below left] {$E$};
            \node at (F) [below right] {$F$};
            \node at (G) [below right] {$G$};
            \node at (H) [left] {$H$};
            \end{tikzpicture}
            \subcaption*{second polygon}
    \end{subfigure}
\end{figure}

On a 2x2x2 lattice, there are 12 such hexagons that must 
have the same constants, denoted as $\alpha$, in order to express 
the statistical sum as an integral over Grassmann variables. 
Computational calculations were performed to determine whether 
there exists an action for which all 12 hexagons have the same 
constants, $\alpha$. Additionally, it was assumed that the action 
in three dimensions should yield the same constants for two-dimensional 
polygons, which was verified by examining the weights of squares. 
For those operations that satisfied the additional assumption, the 
weights of the 12 hexagons were listed. It turned out that no choice 
of the action resulted in the same constants for all hexagons. 
However, upon closer examination of the results, it was found that 
for each action, exactly 6 hexagons had a weight of +1, while 
the remaining 6 had a weight of -1.

Based on these results and analyzing the weights of individual 
hexagons, it was possible to establish that if a hexagon has a 
weight of $\alpha$, its mirror image will have a weight of $-\alpha$. 
For example, regardless of the weight of the first polygon in the above 
diagram, the second polygon, which is its mirror image with respect to 
the horizontal plane, will have the opposite sign.

This observation was proven rigorously assuming that the weight of 
a square is positive. A separate proof is required in the case where 
the sign of the square is negative. Both cases are discussed below.

Case 1: $sgn(\alpha(r_1, r_2, r_3)) = +1$ for $r_1$, $r_2$, $r_3$ 
corresponding to the simplest polygon, i.e., a square with a side 
length of 1 in a chosen plane.

Two polygons contribute $\alpha_1$ and $\alpha_2$, which we express 
using an auxiliary function (we list the terms for the first polygon 
starting from point E, and for the second polygon starting from point H).

\begin{align*}
    \displaystyle
    \alpha_1 &= -F_3(X',Y)F_3(Y',X) F_3(X',Z)F_3(Z',Y')F_3(Y,Z')F_3(Z,X')\\
        &= -F_3(X',Y)F_3(Y',X) \cdot F_3(X',Z)F_3(Z,X') \cdot F_3(Z',Y')F_3(Y,Z')\\
    \alpha_2 &= -F_3(X',Y)F_3(Y',X)  F_3(X',Z')F_3(Z,Y') F_3(Y,Z)F_3(Z',X') \\
        &= -F_3(X',Y)F_3(Y',X) \cdot F_3(X',Z')F_3(Z',X') \cdot F_3(Z,Y')F_3(Y,Z)
\end{align*}

The words are grouped into three pairs, and now we will compare these pairs 
with each other.
The first pair is the same in both expressions. The second pair differs 
by replacing $Z$ with $Z'$, because we obtained new variables through the 
rotation, $Z$ and $Z'$ must have the same type. Consequently, 
they belong to different groups and because $X'$ has a fixed group, either 
$Z$ or $Z'$ should belong to the same group as $X'$. We know that swapping 
the arguments of a function changes the sign only if the arguments belong to 
the same group. Thus, in one of these products, we have two terms with the same 
sign, and in the other, we have terms with opposite signs. We don't know which 
pair has which sign, but we can certainly state that these two pairs have 
opposite signs, that is, $sgn(F_3(X',Z)F_3(Z,X')) = -sgn(F_3(X',Z')F_3(Z',X'))$.
Let's analyze the last pairs. Notice that their product satisfies 
$F_3(Z',Y')F_3(Y,Z') \cdot F_3(Z,Y')F_3(Y,Z) = F_3(Z',Y')F_3(Y,Z') \cdot (-F_3(Y',Z)F_3(Z,Y))$. 
We obtain this from similar reasoning as before. The terms $Z,Y'$ or $Z,Y$ 
belong to the same group, while the remaining pair of terms belongs to 
different groups. Swapping the arguments in one of the terms, either $F_3(Y',Z)$ or 
$F_3(Z,Y)$, changes the sign, while it remains the same in the other term. We don't 
know the behavior of each term, but their product certainly changes the sign. If we 
change the order of the terms, we get 
$F_3(Z',Y')F_3(Y,Z') \cdot F_3(Z,Y')F_3(Y,Z) = -F_3(Z',Y')F_3(Y,Z') \cdot F_3(Z,Y)F_3(Y',Z)$, 
which is the exact form of the expression for $\alpha$ for the square. 
Here, we use our assumption $sgn(\alpha) = +1$.
Therefore, $sgn(F_3(Z',Y')F_3(Y,Z') \cdot F_3(Z,Y')F_3(Y,Z)) = +1$, 
which implies that $sgn(F_3(Z',Y')F_3(Y,Z')) = sgn(F_3(Z,Y')F_3(Y,Z))$. 
Comparing the terms for $\alpha_1$ and $\alpha_2$ gives us the relationship 
$sgn(\alpha_1) = -sgn(\alpha_2)$. Of course, since the range of values for 
$F_3$ does not include 0, $\alpha$ is different from 0. Therefore, the condition 
$sgn(\alpha_1) = -sgn(\alpha_2)$ contradicts the assumption that the constant 
alpha depends only on $r_1$, $r_2$, $r_3$.

Case 2: $sgn(\alpha(r_1, r_2, r_3)) = -1$ for $r_1$, $r_2$, $r_3$ 
corresponding to the simplest polygon, namely a square with a side length of 1, 
located in a chosen plane.

Let's consider two configurations on a piece of the distinguished plane.

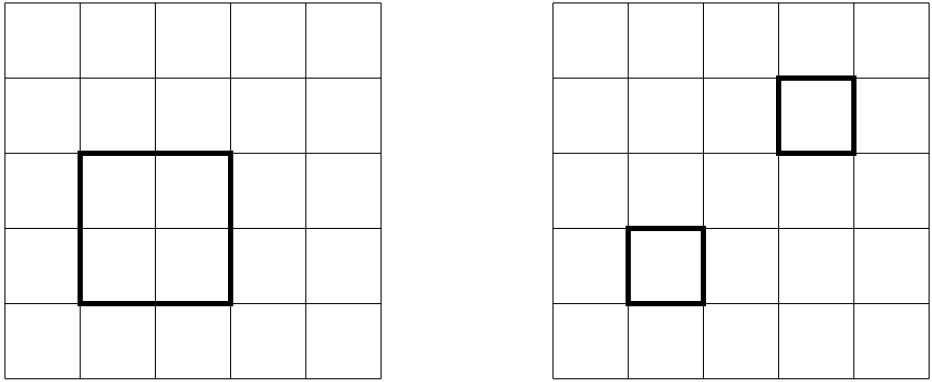
\begin{figure}[H]
    \centering
    \begin{tikzpicture}
        \def\cellsize{1cm}
        % Draw the horizontal lines
        \foreach \y in {0,...,5} {
            \draw (0,\y*\cellsize) -- (5*\cellsize,\y*\cellsize);
        }
        % Draw the vertical lines
        \foreach \x in {0,...,5} {
            \draw (\x*\cellsize,0) -- (\x*\cellsize,5*\cellsize);
        }
        \draw[line width = 2] (1,1) -- (1,3) -- (3,3) -- (3,1) --cycle;
    \end{tikzpicture}
    \hspace*{2 cm}
    \begin{tikzpicture}
        \def\cellsize{1cm}
        % Draw the horizontal lines
        \foreach \y in {0,...,5} {
            \draw (0,\y*\cellsize) -- (5*\cellsize,\y*\cellsize);
        }
        % Draw the vertical lines
        \foreach \x in {0,...,5} {
            \draw (\x*\cellsize,0) -- (\x*\cellsize,5*\cellsize);
        }
        \draw[line width = 2] (1,1) -- (1,2) -- (2,2) -- (2,1) --cycle;
        \draw[line width = 2] (3,3) -- (3,4) -- (4,4) -- (4,3) --cycle;
    \end{tikzpicture}
    \caption*{Two considered configuration}
\end{figure}

Both configurations have 4 edges in each direction and 8 
points belonging to polygons through which the edges pass, 
consequently they have the same number of empty sites. The constants 
$\alpha$ for both configurations can be calculated using the function $F_3$.

\begin{align*}
    \displaystyle
    \alpha_1 &= -F_3(Y',X')F_3(X,X')F_3(X,Y')F_3(Y,Y')F_3(Y,X)F_3(X',X)F_3(X,Y)F_3(Y',Y)\\
        &= -F_3(Y',X')F_3(X,Y')F_3(Y,X)F_3(X,Y) \cdot F_3(X,X')F_3(X',X) \cdot F_3(Y,Y')F_3(Y',Y)\\
    \alpha_2 &= -F_3(Y',X')F_3(X,Y')F_3(Y,X)F_3(X,Y) \cdot -F_3(Y',X')F_3(X,Y')F_3(Y,X)F_3(X,Y)
\end{align*}

Let us examine the signs of both constants. In $\alpha_2$, we have 
the product of two terms, each of which is an expression for the unit 
square with a sign of -1. The product of two such expressions gives 
a sign of +1, that is, $sgn(\alpha_2) = +1$. In the first term, $\alpha_1$, 
we again recognize an expression for the constant of the unit square, 
which has a sign of -1. The remaining two expressions are products of 
two $F_3$ with reversed arguments. Knowing that both pairs X, X' and Y, Y' 
have elements from different groups, we conclude that reversing the arguments 
does not change the sign. Therefore, the signs of both $F_3(X,X')F_3(X',X)$ 
and $F_3(Y,Y')F_3(Y',Y)$ are positive, which gives us $sgn(\alpha_1) = -1$, 
once again contradicting the thesis.

This proof demonstrates the failure of Samuel's method even on 
the smallest lattices. However, it turns out that this can be shown much 
more simply using a slightly larger lattice. The idea is based on creating 
two polygons that not only have the same number of edges in each direction 
but are also composed of the same corners. The only difference between the 
polygons is the directions in which certain corners are traversed, as changing 
the direction can introduce an additional sign of '-'. Clever construction of 
the polygons allows for a much simpler proof. Additionally, the previous 
proof required the constants $a_i$ to be real numbers. However, this proof 
does not rely on this assumption and holds true even for complex numbers.

Consider the following two polygons:

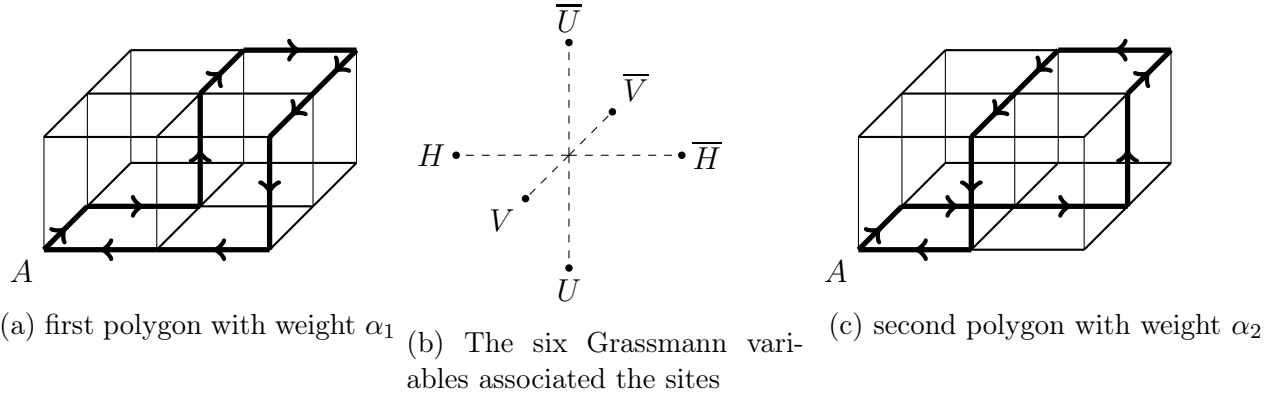
\begin{figure}[H]
    \begin{subfigure}[h]{0.3\linewidth}
        \begin{tikzpicture}[scale=1.5, decoration={markings, mark=at position 0.5 with {\arrow{>}}}    ]
            % draw x lines
            \draw[thick] (0,0,0) -- (2,0,0);
            \draw[thick] (0,0,1) -- (2,0,1);
            \draw[thick] (0,0,2) -- (2,0,2);
            \draw[thick] (0,1,0) -- (2,1,0);
            \draw[thick] (0,1,1) -- (2,1,1);
            \draw[thick] (0,1,2) -- (2,1,2);
            % draw y lines
            \draw[thick] (0,0,0) -- (0,0,2);
            \draw[thick] (1,0,0) -- (1,0,2);
            \draw[thick] (2,0,0) -- (2,0,2);
            \draw[thick] (0,1,0) -- (0,1,2);
            \draw[thick] (1,1,0) -- (1,1,2);
            \draw[thick] (2,1,0) -- (2,1,2);
            % draw z lines
            \foreach \x in {0,1,2}
                \foreach \y in {0,1,2} {
                    \draw (\x,0,\y) -- (\x,1,\y);
                }
            \coordinate (a) at (0,0,2);
            \coordinate (b) at (0,0,1);
            \coordinate (c) at (1,0,1);
            \coordinate (d) at (1,1,1);
            \coordinate (e) at (1,1,0);
            \coordinate (f) at (2,1,0);
            \coordinate (g) at (2,1,1);
            \coordinate (h) at (2,1,2);
            \coordinate (i) at (2,0,2);
            \coordinate (j) at (1,0,2);
            \foreach \start/\end in {a/b,b/c,c/d,d/e,e/f,f/g,g/h,h/i,i/j,j/a}
                \draw[postaction={decorate},line width=2pt]  (\start) -- (\end);
            \node at (0,0,2) [below left] {$A$};
            \end{tikzpicture}
            \subcaption{first polygon with weight $\alpha_1$}
    \end{subfigure}
    \begin{subfigure}[h]{0.3\linewidth}
        \begin{tikzpicture}[scale = 1.5]
            \coordinate (origin) at (0, 0, 0);
            \coordinate (X1) at (1, 0, 0);
            \coordinate (X2) at (-1, 0, 0);
            \coordinate (Y1) at (0, 0, 1);
            \coordinate (Y2) at (0, 0, -1);
            \coordinate (Z1) at (0, 1, 0);
            \coordinate (Z2) at (0, -1, 0);
            \foreach \pt in {X1, X2, Y1, Y2, Z1, Z2}
                \draw[dashed] (origin) -- (\pt);
            \foreach \pt in {X1, X2, Y1, Y2, Z1, Z2}
                \draw (\pt) node[circle, fill, inner sep=1pt] {};
            \node at (X1) [right] {$\overline{H}$};
            \node at (X2) [left] {$H$};
            \node at (Y1) [below left] {$V$};
            \node at (Y2) [above right] {$\overline{V}$};
            \node at (Z1) [above] {$\overline{U}$};
            \node at (Z2) [below] {$U$};
        \end{tikzpicture}
        \subcaption{The six Grassmann variables associated the sites}
    \end{subfigure}
    \begin{subfigure}[h]{0.35\linewidth}
        \begin{tikzpicture}[scale=1.5, decoration={markings, mark=at position 0.5 with {\arrow{>}}}    ]
            % draw x lines
            \draw[thick] (0,0,0) -- (2,0,0);
            \draw[thick] (0,0,1) -- (2,0,1);
            \draw[thick] (0,0,2) -- (2,0,2);
            \draw[thick] (0,1,0) -- (2,1,0);
            \draw[thick] (0,1,1) -- (2,1,1);
            \draw[thick] (0,1,2) -- (2,1,2);
            % draw y lines
            \draw[thick] (0,0,0) -- (0,0,2);
            \draw[thick] (1,0,0) -- (1,0,2);
            \draw[thick] (2,0,0) -- (2,0,2);
            \draw[thick] (0,1,0) -- (0,1,2);
            \draw[thick] (1,1,0) -- (1,1,2);
            \draw[thick] (2,1,0) -- (2,1,2);
            % draw z lines
            \foreach \x in {0,1,2}
                \foreach \y in {0,1,2} {
                    \draw (\x,0,\y) -- (\x,1,\y);
                }
            \coordinate (a) at (0,0,2);
            \coordinate (b) at (0,0,1);
            \coordinate (c) at (1,0,1);
            \coordinate (d) at (2,0,1);
            \coordinate (e) at (2,1,1);
            \coordinate (f) at (2,1,0);
            \coordinate (g) at (1,1,0);
            \coordinate (h) at (1,1,1);
            \coordinate (i) at (1,1,2);
            \coordinate (j) at (1,0,2);
            \foreach \start/\end in {a/b,b/c,c/d,d/e,e/f,f/g,g/h,h/i,i/j,j/a}
                \draw[postaction={decorate},line width=2pt]  (\start) -- (\end);
            \node at (0,0,2) [below left] {$A$};
            \end{tikzpicture}
            \subcaption{second polygon with weight $\alpha_2$}
    \end{subfigure}
\end{figure}

Both polygons have 10 sites and, respectively, 4, 4, and 2 edges in the H, V, 
and U directions, so according to the assumption, both should have the same weights.
Grassmann variables at each site are oriented as shown in the figure.
Let us now express the weights of both polygons, $\alpha_1$ and $\alpha_2$, in 
terms of $F_3$. Both of these weights are written starting from the term for 
site A and going clockwise along the polygon.

\begin{align*}
    \alpha_1 = &\underline{F_3(\overline{H}, \overline{V}) F_3(V, \overline{H})} 
    \cdot \dashuline{F_3(H, \overline{U}) F_3(U, \overline{V})} \cdot \\ \cdot
    &F_3(V, \overline{H}) F_3(H, V) \cdot \dotuline{F_3(\overline{V}, V)} \cdot 
    \uwave{F_3(\overline{V}, U) F_3(\overline{U}, H)} \cdot F_3(\overline{H}, H) \\
    \alpha_2 = &\underline{F_3(\overline{H}, \overline{V}) F_3(V, \overline{H})} 
    \cdot F_3(H, \overline{H}) \cdot
    \dashuline{F_3(H, \overline{U}) F_3(U, \overline{V})} \cdot
    \\ \cdot 
    &F_3(V, H) F_3(\overline{H}, V) \cdot \dotuline{F_3(\overline{V}, V)} \cdot 
    \uwave{F_3(\overline{V}, U) F_3(\overline{U}, H)}
\end{align*}

The same terms are underlined in the equations. Therefore, 
if $\alpha_1$ and $\alpha_2$ are to be equal to each other, 
following condition must be satisfied: 
$F_3(H, \overline{H}) F_3(V, H) F_3(\overline{H}, V) =
F_3(V, \overline{H}) F_3(H, V)F_3(\overline{H}, H)$.
Since $H$ and $\overline{H}$ belong to different groups, 
we have $F_3(H, \overline{H}) = F_3(\overline{H}, H)$; for 
the same reason, $F_3(V, \overline{H}) = F_3(\overline{H}, V)$.
However, since H and V are in the same group, we have $F_3(H, V) = -F_3(V, H)$, 
which ultimately gives us $F_3(H, \overline{H}) F_3(V, H) F_3(\overline{H}, V) = -
F_3(V, \overline{H}) F_3(H, V)F_3(\overline{H}, H)$, contradicting the thesis.

%%%%%%%%%%%%%%%%%%%%%%%%%%%%%%%%%%%%
\section{Summary and conclusions}
\label{sec:Wnioski}

In our paper, we examined the possibility of computing the partition function for the 3D Ising model as a Grassmann integral over bilinear action. 
This possibility is well known in 2D, but some mentions of non-extensibility of such approach to 3D were rare and loose. We wanted to take closer look to this opportunity and understand reason of this presumed non-extensibility. 
We have examined the most general expression for a bilinear action in Grassmann variables, integral of which would give the polygonal form (high-temperature expansion) of the partition function.
However, it turned out that such an expression {\it does not} replicate the partition function:
The number of polygons is correct, but some of them enter with a {\it negative sign}. 
In other words, it has been shown that it is not possible to derive the expression for the partition function of the 3D Ising model using the translationally invariant bilinear action.
It can be interpreted as manifestation of the fact that mathematics behind 2D 
and 3D Ising model are essentially different: the 2D Ising model can be viewed as a
system of non-interacting fermions, whereas in 3D, the fermionic picture inevitably leads
to interacting fermions \cite{Samuel3}, \cite{Nojima}. 

The technique of Grassmann integrals is intimately connected with {\it dimers}. 
For dimers on 2D lattices, there are lot of theorems allow for effective  computations of dimer coverings \cite{Hurst&Green}, \cite{Aizenman&Warzel}, \cite{Hurst}, \cite{Kasteleyn}. 
As far as the authors know, there are no analogous theorems in 3D. Some attempts in this
direction were mentioned in a few papers, for instance \cite{Zecchina1}, \cite{Zecchina2}, \cite{Plechko2}, but they did not lead to a general 3D case. 
It would be very interesting to relate Grassmann integrals in 3D to dimers on 3D lattices.


\begin{thebibliography}{10}
\bibitem{Lenz} W. Lenz, {\it Z. Phys.} \textbf{21}, 613 (1920)
\bibitem{Ising} E. Ising, {\it Z. Phys.} \textbf{31}, 253 (1925)
\bibitem{Onsager} L. Onsager, {\it Phys. Rev.} \textbf{65}, 117 (1944)
\bibitem{Kaufman} B. Kaufman, {\it Phys. Rev.} \textbf{76}, 1232 (1949)
\bibitem{Kac&Ward} M. Kac and J. C. Ward, {\it Phys. Rev.} \textbf{88}, 1332 (1952).
\bibitem{LSM}  T. D. Schulz,  D. C. Mattis, and E. H. Lieb,  {\it Rev. Mod. Phys.}\textbf{36}, 856 (1964).
\bibitem{Vdv} N. V. Vdovichenko,  {\it J. Exptl. Theor. Phys. (USSR)} \textbf{47}, 715 (1964). Translation:  {\it Soviet Physics JETP} \textbf{20}, 477 (1965)
\bibitem{Baxter&Enting} R. J. Baxter and I. G. Enting, {\it J. Phys. A: Math. Gen.}
 \textbf{11}, 2463 (1978)
\bibitem{Rychkov} S. Rychkov,    {\it Comptes Rendus. Physique}, Tome \textbf{21} (2020) no. 2, pp. 185-198
\bibitem{Samuel1} S. Samuel, {\it J. Math. Phys.} \textbf{21}, 2806 (1980)
\bibitem{Samuel3} S. Samuel, {\it J. Math. Phys.} \textbf{21} 2820 (1980)
\bibitem{Zhang} D. Zhang, (2021) { \sf{arXiv:2110.11233v1} }
\bibitem{Zhang-com} J. Perk, (2022){ \sf{ arXiv:2202.03136v4}}
\bibitem{Zhang-com-com} D. Zhang, (2023) { \sf{arXiv:2302.10139}}
\bibitem{Viswanathan} G. Viswanathan, M. A. Portillo, E. Raposo, M. da Luz, {\it Entropy} \textbf{24}, 1665 (2022)
\bibitem{Giuliani} A. Giuliani, { \sf{arXiv:cond-mat/0412176v1 [cond-mat.stat-mech]}}
\bibitem{Hurst&Green} C. A. Hurst and H. S. Green, {\it J. Chem. Phys.} \textbf{33}, 1059 (1960)
%\bibitem{Pearson}R. B. Pearson, {\it J. Phys.} \textbf{26}, 2463 (1978)
%\bibitem{Yang&Lee}C. N. Yang and T. D. Lee, {\it Phys. Rev. B} \textbf{87}, 6285 (1982);
%C. N. Yang and T. D. Lee, {\it Phys. Rev.} \textbf{87}, 410 (1952);
%\bibitem{McKenzie} S. McKenzie, (1982). Derivation of High Temperature Series Expansions: Ising Model. In: Levy, M., Le Guillou, J. C., Zinn-Justin, J. (eds.): Phase Transitions Cargese 1980. Springer, Boston, MA.
\bibitem{Wei} R. Q. Wei, (2018) { \sf{arXiv:1805.01366v3}}
\bibitem{Aizenman&Warzel}M. Aizenman and S. Warzel, {\it J. Stat Phys} \textbf{173}, 1755-1778 (2018)
\bibitem{Zecchina1}T. Regge and R. Zecchina, {\it J. Phys. A: Math. Gen.} \textbf{33}, 741-761 (2000)
\bibitem{Zecchina2}T. Regge and R. Zecchina, (1996) { \sf{arXiv:cond-mat/9603072v1}}
\bibitem{Wegner}F. Wegner, {\it J. Math. Phys.} \textbf{12}, 2259 (1971)
\bibitem{Hurst} C. A. Hurst, {\it J. Math. Phys.} \textbf{5}, 90 (1964)
\bibitem{Kasteleyn} P. W. Kasteleyn, Graph theory and crystal physics. In:  F. Harary (ed.): Graph Theory
and Theoretical Physics,  pp. 43-110, Academic Press, London 1967.
\bibitem{Nojima} K. Nojima, {\it Int J. Mod. Phys. B} \textbf{12}, 1995 (1998)
%\bibitem{Plechko1} V. N. Plechko, {\it Phys. Lett. A} \textbf{239}, 289 (1998)
\bibitem{Plechko2} V. N. Plechko, {\it J. Phys. Studies} Vol. \textbf{1}, No. \textbf{4} 554 (1997)
%\bibitem{Plechko3} V. N. Plechko, {\it Physica A} Vol. \textbf{152}, 51 (1988)
%\bibitem{Plechko4} V. N. Plechko, (2004) { \sf{arXiv:math-ph/0411084v1}}






\end{thebibliography}
\end{document}